\documentclass[pdflatex,sn-mathphys-num]{sn-jnl}

\usepackage{graphicx}%
\usepackage{multirow}%
\usepackage{amsmath,amssymb,amsfonts}%
\usepackage{amsthm}%
\usepackage{mathrsfs}%
\usepackage[title]{appendix}%
\usepackage{xcolor}%
\usepackage{textcomp}%
\usepackage{manyfoot}%
\usepackage{booktabs}%
\usepackage{algorithm}%
\usepackage{algorithmicx}%
\usepackage{algpseudocode}%
\usepackage{listings}%
\usepackage{xcolor}
\usepackage{bbding}

\theoremstyle{thmstyleone}%
\theoremstyle{thmstyletwo}%

\theoremstyle{thmstylethree}%

\begin{document}

\title[Article Title]{On the Quantization of the Motion of a Liquid}


\author{\fnm{Nadine} \sur{Suzan Cetin}}\email{nadine-suzan.cetin@matfyz.cuni.cz}



\affil{\orgdiv{Faculty of Mathematics and Physics, Mathematical Institute}, \orgname{Charles University}, \orgaddress{\street{Sokolovská 49/83}, \city{Prague}, \postcode{12000}, \country{Czech Republic}}}




\abstract{For building up a theory of superfluid Helium-4, Lev Landau ingeniously unified the principles of quantum mechanics with the principles of hydrodynamics. By introducing a velocity operator he was able to derive a quantum analogue of the mass density flux equation and a quantum Euler equation. The article shows that it is possible to formulate a consistent \textit{quantum} theory of the ideal \textit{liquid} based on Landau's Quantum Hydrodynamics. This means that it is added the quantum analogue of the momentum density flux equation, the energy density flux equation, the entropy density flux equation, and a quantum vorticity equation. These are all manifestly in agreement with the classical limit of the flux equations of classical ideal liquids. It is also shown that the various aspects that have been criticised about Landau's Quantum Hydrodynamics - e.g. the application of the inverse of the mass density operator - are not justified. Such a consistent quantum theory of an ideal liquid may therefore provide a starting point of a complementary approach in the theoretical research of superfluid Helium 4.}


\maketitle

\section{Introduction}
Landaus theory of superfluid Helium 4 begins with a first attempt to bring together the principles of quantum mechanics with the principles of hydrodynamics \cite{Landau41}. In this seminal work L. Landau introduced the momentum density operator $\hat{j}_{a}(\mathbf{r}_{1})$, the mass density operator $\hat{\rho}(\mathbf{r}_{1})$ and the \textit{velocity operator} $\hat{v}_{a}(\mathbf{r}_{1})$ in order to describe the liquid \textit{as a quantum liquid} as

\begin{equation} \label{operatorsLQHD}
    \begin{aligned}
        \hat{j}_{a}(\mathbf{r}_{1})&=\frac{1}{2}\left(\hat{p}_{a}\delta(\mathbf{r}-\mathbf{r}_{1})+\delta(\mathbf{r}-\mathbf{r}_{1})\hat{p}_{a}\right)\\
        \hat{\rho}(\mathbf{r}_{1})&=m\delta(\mathbf{r}-\mathbf{r}_{1})\\
        \hat{v}_{a}(\mathbf{r}_{1})&=\frac{1}{2}\left(\frac{1}{\hat{\rho}(\mathbf{r}_{1})}\hat{j}_{a}(\mathbf{r}_{1})+\hat{j}_{a}(\mathbf{r}_{1})\frac{1}{\hat{\rho}(\mathbf{r}_{1})}\right)\\
        \hat p_a&= \frac{\hbar}{i}\frac{\partial}{\partial r_a}
    \end{aligned}
\end{equation}

$\mathbf r$ is the position of the particle, $\mathbf r_1$ is an arbitrary point in the liquid, and  
 
\begin{equation} \label{operatorsconjugated}
    \begin{aligned}
        \hat{j}^\dagger_{a}(\mathbf{r}_{1})&=\hat{j}_{a}(\mathbf{r}_{1}) \\
     \hat{v}^\dagger_{a}(\mathbf{r}_{1})&=\hat{v}_{a}(\mathbf{r}_{1})
    \end{aligned}
\end{equation}

The symmetrized, Hermitian Hamiltonian $\hat H$ of the quantum liquid is given by 

\begin{equation}\label{Hamiltonian}
\begin{aligned}
    \hat{H}&=\int d^{3}r_{1}\left(\frac{\hat{v}_{a}(\mathbf{r}_{1})\hat{\rho}(\mathbf{r}_{1})\hat{v}_{a}(\mathbf{r}_{1})}{2}+\hat{\varepsilon}\left(\hat{\rho}(\mathbf{r}_{1})\right)\right)\\
 \hat{H}^\dagger&=\hat{H}
\end{aligned}
\end{equation}

where $\hat{\varepsilon}\left(\hat{\rho}(\mathbf{r}_{1})\right)$ is the internal energy density operator \footnote{In the original work there is an additional $\hat \rho (\mathbf r_1)$ multiplied with $\hat{\varepsilon}\left(\hat{\rho}(\mathbf{r}_{1})\right)$. This must be a misprint, see also \cite{DzyaloshinskiVolovik80, KhalatnikovBook}}.

Landau derived the commutation relation between the momentum density $\hat{j}_{a}(\mathbf{r}_{1})$ and the mass density $\hat{\rho}(\mathbf{r}_{2})$ as

\begin{equation} \label{j-rho}
  \left[\hat{j}_{a}(\mathbf{r}_{1}),\hat{\rho}(\mathbf{r}_{2})\right]=\frac{\hbar}{i}\hat{\rho}(\mathbf{r}_{1})
  \begin{array}{c}
\frac{\partial\delta\left(\mathbf{r_1}-\mathbf{r}_{2}\right)}{\partial r_{a}}
\end{array}
\end{equation} 

Here, $\frac{\partial\delta\left(\mathbf{r}_{1}-\mathbf{r}_{2}\right)}{\partial r_{a}}$ is the gradient of the delta–distribution which means that the derivative does not act as an operator any more.

The commutation relation between the velocity operator and the mass density operator is then given by

\begin{equation} \label{v-rho}
    \left[\hat{v}_{a}(\mathbf{r}_{1}),\hat{\rho}(\mathbf{r}_{2})\right]=\frac{\hbar}{i}\left(\frac{\partial\delta\left(\mathbf{r}_{1}-\mathbf{r}_{2}\right)}{\partial r_{a}}\right)\hat{1}
\end{equation}

Finally, Landau provided the commutator between the components $a,b\in\{x,y,z\}$ of the velocity operator as

\begin{equation} \label{va-vb}
   \begin{aligned}
       \left[\hat{v}_{a}(\mathbf{r}_{1}),\hat{v}_{b}(\mathbf{r}_{2})\right]&=\frac{\hbar}{i}\delta\left(\mathbf{r}_{2}-\mathbf{r}_{1}\right)\frac{1}{\hat{\rho}(\mathbf{r}_{1})}\left(\mathrm{rot}\,\hat{v}(\mathbf{r}_{1})\right)_{ab}\\
        \left(\mathrm{rot}\,\hat{v}(\mathbf{r}_{1})\right)_{ab}&=\frac{\partial \hat{v}_{b}(\mathbf{r}_{1})}{\text{\ensuremath{\partial r_{a}}}}-\frac{\partial \hat{v}_{a}(\mathbf{r}_{1})}{\text{\ensuremath{\partial r_{b}}}}
   \end{aligned}
\end{equation}

Landau showed that, in the Heisenberg representation of the mass density operator $\hat \rho (\mathbf r_2)$ and the velocity operator $\hat v_b (\mathbf r_2)$, for the related Heisenberg equations of motion, the quantum analogue of the mass balance equation follows as:

\begin{equation} \label{massbalance}
    \begin{aligned}
        \partial_{t}\hat{\rho}(\mathbf{r}_{2})&=\frac{i}{\hbar}\left[\hat{H},\hat{\rho}(\mathbf{r}_{2})\right]=-\mathrm{div}\hat{j}(\mathbf{r}_{2})
    \end{aligned}
\end{equation}

And, too, the quantum analogue of the Euler equation follows as

\begin{equation} \label{Eulerequation}
    \begin{aligned}
      \partial_{t} &\hat{v}_{b}(\mathbf{r}_{2})+\frac{1}{2}\left(\hat{v}_{a}(\mathbf{r}_{2})\left(\frac{\partial\hat{v}_{b}(\mathbf{r}_{2})}{\partial r_{a}}\right)+\left(\frac{\partial\hat{v}_{b}(\mathbf{r}_{2})}{\partial r_{a}}\right)\hat{v}_{a}(\mathbf{r}_{2})\right)\\
      &=-\frac{1}{\hat{\rho}(\mathbf{r}_{2})}\frac{\partial}{\partial r_{b}}\frac{\partial\hat{\varepsilon}(\mathbf{r}_{2})}{\partial\hat{\rho}}
    \end{aligned}
\end{equation}
This will be called Landau Quantum Hydrodynamics (LQHD) in the following.

Landau's ultimate aim was not to describe the liquid as a liquid, he aimed at describing the liquid by its elementary excitations (all sorts of phonons). It may therefore be that he set up such a quantum hydrodynamics for two other reasons: first, simply because liquid Helium 4 is a quantum system. Second, to give an argument for the energy gap $\Delta$ in the excitation spectrum of the liquid \cite{Ziman53}. From the commutation relation between the velocity components \eqref{va-vb} it was possible to conclude that, like for the angular momentum in quantum mechanics, there is a minimum excitation energy of the order of $\hbar$ in the rotatory states of the liquid which he called rotons. According to Landau, the energy gap $\Delta$ in the spectrum of superfluid Helium 4 exists as "...between the states of the potential ($\mathrm{curl}\, \mathbf{\hat v} = \hat 0$) and vortex ($\mathrm{curl}\, \mathbf{\hat v} \neq \hat 0$) motions of a quantum liquid there is no continuous transition." \cite{Landau41}. 

Such a theory of a quantum liquid is the first an only one of its kind, and it is quite remarkable that a quantum analogue of the Euler equation can be derived from the algebra of commutators of density operators describing the liquid itself. It has to be emphasized, too, that, by assuming only two ingredients, the Hermiticity of all operators and the existence of a velocity operator, the structure of a theory describing the liquid itself, not its elementary excitations, already shimmers through. Finally, as liquid Helium 4 is, indeed, a quantum liquid, it seems to be very interesting to investigate to which extent LQHD covers the current understanding of superfluid Helium 4 as a depleted Bose--Einstein--Condensate. 

The heart of LQHD is the introduction of the velocity operator $\hat v_a(\mathbf r_1)$. This operator, indeed, links quantum mechanics with hydrodynamics. The velocity operator cannot be derived from the momentum density operator $\hat{j}_{a}(\mathbf{r}_{1})$. Landau postulated the existence of such a velocity operator and constructed it in such a way that it would be Hermitian and, in the classical limit, would give back the classical relation between velocity and momentum.

However, the velocity operator postulated by Landau is one of the aspects of the theory that has been most criticized. F. London, e.g. rejected it: "The concept of the 'velocity field' of classical hydrodynamics has apparently no simple counterpart in quantum hydrodynamics and Landau's operator $\mathbf v( R)$ is merely a formal construction devoid of the physical significance attributed to it. It is rather the current density field $j$, not the velocity field $v$, which is to be considered as the appropriate basic quantity of quantum hydrodynamics". \cite{London45}

Assuming that the momentum current density operator $\hat j_a (\mathbf r_1)$ would, indeed, be the fundamental operator of LQHD, Bierter and Morrison tried to construct the theory by using the mass density operator and the momentum current density operator only \cite{BierterMorrison69}. In their work, they did not express the Hamiltonian \eqref{Hamiltonian} by the momentum density operator but used, instead, another Hamiltonian which no longer has the simple structure and interpretability  as the one suggested by Landau. The equation of motion of the momentum density operator (conservation of momentum density) is derived, however, an unjustified approximation in the density operator is used. The final result does obviously not have the structure of the momentum density flux equation of classical hydrodynamics, and, moreover, the result is not commented.

In \cite{Lee69,KobeCoomer73} LQHD is derived from quantum field theoretical methods applied to a many--Boson system. The rather pessimistic conclusion in \cite{KobeCoomer73}, however, is: "It is difficult to be optimistic about the quantized hydrodynamic approach to the quantum many-body problem, since the mathematical problems appear so great. So many dubious formal manipulations were made with operators whose existence is doubtful, that it is remarkable that we obtained Eqs. (4. 1) and (4. 2), which are the operator form of the equation of continuity and Euler's equation, respectively."

Hence, another aspect that was criticized, and what is meant by the "operators whose existence is doubtful" \cite{KobeCoomer73}, is the usage of the inverse operator $\frac{1}{\hat \rho (\mathbf r_1)}$. 

Due to those apparent formal problems of LQHD and the related mistrust into the theory it was also derived from a variational principle. Thellung and Ziman, independently, used the Clebsch decomposition for the velocity field of the flow and derived from a Lagrangian density, which would give the equations of motion of the fluid, a Hamiltonian density. Knowing the conjugated variables one can canonically quantize the system by imposing on commutation relations between the canonically conjagated variables, and symmetrize the Hamiltonian density. In that way it is possible to derive the commutation relations presented by Landau from a variational principle \cite{Thellung53, Ziman53}. 

The sign in the commutator \eqref{va-vb} has been discussed in the literature to some extent. In \cite{KhalatnikovBook, Ziman53} one finds a positive sign like in Landau's original work. Such a positive sign would, however, not yield the quantum Euler equation. In \cite{VolovickHeliumBook} the commutator \eqref{va-vb} is derived from deriving the Poisson brackets by applying the Correspondance Principle to LQHD. In \cite{KobeCoomer73, Lee69} one finds a negative sign. Another possibility to derive a minus sign for the commutator is by applying the Galilean invariance principle to the operators \eqref{operatorsLQHD} \cite{Garrison70}.
 
This publications aims at putting LQHD to a more solid ground. It is shown that all hydrodynamic equations of motion of an ideal liquid can be derived in a quantum mechanical form. This is done following the method proposed by Landau, hence, in an elementary fashion: by applying the well--known mathematical manipulations of quantum mechanics. Next, the vorticity is introduced into LQHD and it is shown that the circulation is quantised by the nature of the theory. With that, to the quantum analogue of the mass density flux \eqref{massbalance} and the quantum Euler equation \eqref{Eulerequation} derived by Landau, is added the momentum density flux equation, the energy density flux equation, the entropy density flux equation and the balance equation of quantum vorticity.

The negative sign in \eqref{va-vb} is derived afoot, which is, indeed, "more tedious" \cite{Lee69}, but not impossible.

Next it is argued that the velocity operator $\hat v_a(\mathbf r_1)$ is a meaningful, useful and essential ingredient of LQHD, and that inverse operator $\frac{1}{\hat \rho (\mathbf r_1)}$ is not a problem, at all, as it can be handled within the standard manipulations of the theory. The conceptual meaning of LQHD will be established and its relation to the two--fluid model and the Gross--Pitaevskii theory is discussed.

\section{Method: the commutation relations of LQHD} \label{sec:Methods}

All commutation relations of LQHD can be derived by applying the elementary mathematical manipulations of quantum mechanics (evaluating commutators or applying the product rule together with keeping track of the order of operators).  

Following the pedestrian path, the only rule that has to be applied in addition is the replacement of the position $\mathbf r$ in the gradients of the delta--distributions $\frac{\partial \delta (\mathbf r-\mathbf r_2)}{\partial r_a}$ by what can be called "outer operator". This rather formal rule has already been used by Landau. Here, it is equipped with the symbol \textcolor{magenta}{\FiveFlowerOpen} such that each step can be followed carefully. The complementary operation of \textcolor{magenta}{\FiveFlowerOpen} is equipped with the symbol \textcolor{green}{\EightFlowerPetal}. For the derivation of the mass density flux equation and the quantum Euler equation only the operation \textcolor{magenta}{\FiveFlowerOpen} is necessary. Applying the operations \textcolor{magenta}{\FiveFlowerOpen} and \textcolor{green}{\EightFlowerPetal} together yield all the other balance equations \eqref{momentum flux result}, \eqref{energy-flux-result}, \eqref{Heisenberg-s} , and \eqref{Heisenberg-omega2} too, in a completely systematic way. This means they can be derived with the correct minus sign on the right hand side the equation, and the correct number of terms hit by the gradients.

An example of what is meant by that shall be given by explicitly deriving of the commutator between the momentum density operator and the mass density operator:

\begin{equation} \label{examplepink}
\begin{aligned}
[\hat{j}_{a}(\mathbf{r}_{1}),\hat{\rho}(\mathbf{r}_{2})]&=\frac{m}{2}\left(\begin{array}{c}
\left[\hat{p}_{a},\delta\left(\mathbf{r}-\mathbf{r}_{2}\right)\right]\delta\left(\mathbf{r}-\mathbf{r}_{1}\right)\\
+\delta\left(\mathbf{r}-\mathbf{r}_{1}\right)\left[\hat{p}_{a},\delta\left(\mathbf{r}-\mathbf{r}_{2}\right)\right]
\end{array}\right)\\
    &=\frac{\hbar}{i}\begin{array}{c}
\frac{\partial\delta\left(\mathbf{r}-\mathbf{r}_{2}\right)}{\partial r_{a}}m\delta\left(\mathbf{r}-\mathbf{r}_{1}\right)\end{array}\\
&\underset{\longrightarrow}{\text{\textcolor{magenta}{\FiveFlowerOpen}}} \frac{\hbar}{i}\begin{array}{c}
\frac{\partial\delta\left(\mathbf{r}_{1}-\mathbf{r}_{2}\right)}{\partial r_{1a}}m\delta\left(\mathbf{r}-\mathbf{r}_{1}\right)\end{array}\\
&=\frac{\hbar}{i}\begin{array}{c}
\frac{\partial\delta\left(\mathbf{r}_{1}-\mathbf{r}_{2}\right)}{\partial r_{1a}}\hat{\rho}(\mathbf{r}_{1})\end{array}
\end{aligned}
\end{equation}

The "outer operator" in this case is the mass density operator $\hat \rho(\mathbf r_1)$.

In the frame of LQHD expressions like \eqref{examplepink} typically appear under an integral as

\begin{equation} \label{examplepink2}
\begin{aligned}
\int d^3 r_1 \frac{\hbar}{i}\begin{array}{c}
\frac{\partial\delta\left(\mathbf{r}_{1}-\mathbf{r}_{2}\right)}{\partial r_{1a}}\hat{\rho}(\mathbf{r}_{1})\end{array} &\underset{=}{p.I.}-\int d^3 r_1 \frac{\hbar}{i}\begin{array}{c}
\bigr(\frac{\partial \hat{\rho}(\mathbf{r}_{1})}{\partial r_{1a}} \bigl) \delta\left(\mathbf{r}_{1}-\mathbf{r}_{2}\right)\end{array} \\
&\underset{\longrightarrow}{\text{\textcolor{green}{\EightFlowerPetal}}}
-\int d^3 r_1 \frac{\hbar}{i}\begin{array}{c}
\bigr(\frac{\partial \hat{\rho}(\mathbf{r}_{1})}{\partial r_{a}} \bigl) \delta\left(\mathbf{r}_{1}-\mathbf{r}_{2}\right)\end{array}\\
&=-\frac{\partial \hat{\rho}(\mathbf{r}_{1})}{\partial r_{a}}
\end{aligned}
\end{equation}

"p.I." means integration by parts. Beware, however, that the operation \textcolor{green}{\EightFlowerPetal} can only be applied after sorting the terms in such a way that the derivatives hit the correct terms, otherwise one collects too many terms under the divergence:\\

\begin{equation}
\begin{aligned}
    \frac{\partial}{\partial r_{1a}}\left(\hat v_a (\mathbf r_1)\hat \rho (\mathbf r_2)\hat v_a (\mathbf r_1) \right)&=\biggl(\frac{\partial}{\partial r_{1a}}\hat v_a (\mathbf r_1)\biggr)\hat \rho (\mathbf r_2)\hat v_a (\mathbf r_1) + \hat \rho (\mathbf r_2)\hat v_a (\mathbf r_1)\biggl(\frac{\partial}{\partial r_{1a}}\hat v_a (\mathbf r_1)\biggr)\\
    &\underset{\longrightarrow}{\text{\textcolor{green}{\EightFlowerPetal}}}
    \biggl(\frac{\partial}{\partial r_{a}}\hat v_a (\mathbf r_1)\biggr)\hat \rho (\mathbf r_2)\hat v_a (\mathbf r_1) + \hat \rho (\mathbf r_2)\hat v_a (\mathbf r_1)\biggl(\frac{\partial}{\partial r_{a}}\hat v_a (\mathbf r_1)\biggr)
    \end{aligned}
\end{equation}

Hence, the operation \textcolor{green}{\EightFlowerPetal} first makes sure that only the the variables are hit by the gradient which are necessary to correctly balance the left hand side (temporal derivative), and second, it reverses the operation \textcolor{magenta}{\FiveFlowerOpen}. It may very well be that there is either an elementary way to proof the correctness of the operations \textcolor{magenta}{\FiveFlowerOpen} and \textcolor{green}{\EightFlowerPetal} (it seems they are somehow related to the product rule), or to find a mathematical foundation within the frame of distribution theory. Landau himself did not change the derivative $\frac{\partial}{\partial r}$, he only replaced $\mathbf r$ in the gradient of the delta--distribution as $\frac{\partial \delta (\mathbf r-\mathbf r_1)}{\partial r_a}\hat \rho (\mathbf r_2)\rightarrow \frac{\partial \delta (\mathbf r_2-\mathbf r_1)}{\partial r_a}\hat \rho (\mathbf r_2)$, see \eqref{j-rho} and \eqref{v-rho}.

The commutation relation between the components of the momentum density operator can be shown to be 
\begin{equation} \label{ja-jb}    \left[\hat{j}_{a}\left(\mathbf{r}_{1}\right),\hat{j}_{b}\left(\mathbf{r}_{2}\right)\right]=-\frac{\hbar}{i}\delta\left(\mathbf{\mathbf{r}_{2}}-\mathbf{r}_{1}\right)\left(\mathrm{rot}\,\hat{j}\left(\mathbf{r}_{1}\right)\right)_{ab}
\end{equation}

Next, much more intricate, but still elementary, it can be shown that the commutation relation between the velocity operator components is given by 

\begin{equation} \label{va-vb correct}    \left[\hat{v}_{a}\left(\mathbf{r}_{1}\right),\hat{v}_{b}\left(\mathbf{r}_{2}\right)\right]=-\frac{\hbar}{i}\delta\left(\mathbf{\mathbf{r}_{2}}-\mathbf{r}_{1}\right) \frac{1}{\hat\rho(\mathbf r_1)}\left(\mathrm{rot}\,\hat{v}\left(\mathbf{r}_{1}\right)\right)_{ab}
\end{equation}

Therefore, by going afoot, one can show that there is a misprint in Landaus original work, where \eqref{va-vb correct} comes positive.  The minus sign, however, is essential for all the balance equations.

The commutation relation between the components of the momentum density operator and the velocity operator can be shown to be

\begin{equation} \label{va-jb}
    \left[\hat{v}_{a}(\mathbf{r}_{1}),\hat{j}_{b}(\mathbf{r}_{2})\right]=\frac{\hbar}{i}\frac{1}{2}\left(\begin{array}{c}
-\delta\left(\mathbf{r}_{2}-\mathbf{r}_{1}\right)\left(\frac{1}{\hat{\rho}(\mathbf{r}_{1})}\left(\mathrm{rot}\hat{v}(\mathbf{r}_{1})\right)_{ab}\hat{\rho}(\mathbf{r}_{2})+\hat{\rho}(\mathbf{r}_{2})\frac{1}{\hat{\rho}(\mathbf{r}_{1})}\left(\mathrm{rot}\hat{v}(\mathbf{r}_{1})\right)_{ab}\right)\\
\left(\frac{\partial\delta\left(\mathbf{r}_{1}-\mathbf{r}_{2}\right)}{\partial r_{1a}}\right)\hat{1}\hat{v}_{b}(\mathbf{r}_{2})+\hat{v}_{b}(\mathbf{r}_{2})\left(\frac{\partial\delta\left(\mathbf{r}_{1}-\mathbf{r}_{2}\right)}{\partial r_{1a}}\right)\hat{1}
\end{array}\right)
\end{equation}

For any function $\hat f\left(\hat\rho(\mathbf r_1) \right)$ one can show that

\begin{equation}\label{f(rho)-vb}
    \left[\hat{f}(\hat{\rho}(\mathbf{r}_{1})),\hat{v}_{b}(\mathbf{r}_{2})\right]=-\frac{\hbar}{i}\left(\frac{\partial\delta(\mathbf{r}_{2}-\mathbf{r}_{1})}{\partial r_{2b}}\right)\frac{\partial\hat{f}(\hat{\rho}(\mathbf{r}_{1}))}{\partial\hat{\rho}}
\end{equation}

which is necessary to know in the case that $\hat f(\mathbf r_2)=\hat \varepsilon(\hat\rho(\mathbf r_2))$.

Then it can be shown that for any function $\hat f(\mathbf r_2)$ there holds 

\begin{equation} \label{rotv-f}
    \left[\left(\mathrm{rot}\hat{v}(\mathbf{r}_{1})\right)_{a},\hat{f}(\mathbf{r}_{2})\right]=\left[\varepsilon_{cba}\frac{\partial}{\partial r_{c}}\hat{v}_{b}(\mathbf{r}_{1}),\hat{f}(\mathbf{r}_{2})\right]=\hat 0
\end{equation}

Which turns out to be important for commutators like $\left[\mathrm{rot}\hat{v}(\mathbf{r}_{1}),\hat \rho(\mathbf r_2) \right]$.

Also of importance is the commutator

\begin{equation} \label{v,omega}
\left[\hat{v}_{a}(\mathbf{r}_{1}),\hat{\omega}_{c}(\mathbf{r}_{2})\right]=-\frac{\hbar}{i}\delta(\mathbf{r}_{1}-\mathbf{r}_{2})\varepsilon_{bdc}\frac{\partial}{\partial r}_{b}\left(\frac{1}{\hat{\rho}(\mathbf{r}_{1})}\left(\frac{\partial v_{d}(\mathbf{r}_{1})}{\partial r_{a}}-\frac{\partial v_{a}(\mathbf{r}_{1})}{\partial r_{d}}\right)\right)
\end{equation}

as it has to be used for deriving \eqref{Heisenberg-omega}. 

Finally, for the sake of completeness,

\begin{equation} \label{rotv-rotv}
 \left[\left(\mathrm{rot}\hat{v}(\mathbf{r}_{1})\right)_{ab},\left(\mathrm{rot}\hat{v}(\mathbf{r}_{2})\right)_{cd}\right]=\hat{0}
\end{equation}

\section{Results}
\subsection*{Momentum Density Flux Equation}
In the Heisenberg representation the momentum density flux equation is given by

\begin{equation} \label{Heisenberg-j}
    \partial_{t}\hat{j}_{b}(\mathbf{r}_{2})=\frac{i}{\hbar} \left[\hat{H},\hat{j}_{b}(\mathbf{r}_{2})\right]
\end{equation}

Evaluating the commutator yields

\begin{equation}\label{momentum flux result}
    \partial_t\hat j_b(\mathbf r_2)=-\frac{\partial\mathfrak{\hat{p}}(\mathbf r_2)}{\partial r_{b}}-\frac{1}{2}\left(\begin{array}{c}
\hat{v}_{a}(\mathbf{r}_{2})\hat{\rho}(\mathbf{r}_{2})\left[\frac{\partial}{\partial r_{a}}\hat{v}_{b}(\mathbf{r}_{2})\right]+\left[\frac{\partial}{\partial r_{a}}\hat{v}_{b}(\mathbf{r}_{2})\right]\hat{v}_{a}(\mathbf{r}_{2})\hat{\rho}(\mathbf{r}_{2})\\
+\left(\frac{\partial}{\partial r_{a}}\hat{v}_{a}(\mathbf{r}_{2})\right)\hat{\rho}(\mathbf{r}_{2})\hat{v}_{b}(\mathbf{r}_{2})+\hat{v}_{b}(\mathbf{r}_{2})\left(\frac{\partial}{\partial r_{a}}\hat{v}_{a}(\mathbf{r}_{2})\right)\hat{\rho}(\mathbf{r}_{2})\\
+\hat{v}_{a}(\mathbf{r}_{2})\left(\frac{\partial}{\partial r_{a}}\hat{\rho}(\mathbf{r}_{2})\right)\hat{v}_{b}(\mathbf{r}_{2})+\hat{v}_{b}(\mathbf{r}_{2})\hat{v}_{a}(\mathbf{r}_{2})\left(\frac{\partial}{\partial r_{a}}\hat{\rho}(\mathbf{r}_{2})\right)
\end{array}\right)
\end{equation}

Here use has been made of the thermodynamic relation $\frac{\partial}{\partial r_{b}}\frac{\partial\hat{\varepsilon}(\hat{\rho}(\mathbf{r}_{2}))}{\partial\hat{\rho}}=\frac{1}{\hat{\rho}(\mathbf{r}_{2})}\frac{\partial\mathfrak{\hat{p}}(\mathbf r_2)}{\partial r_{b}}$ which defines the pressure $\mathfrak{\hat{p}}(\mathbf r_2)$ of the liquid.

Eq. \eqref{momentum flux result} is the symmetrized, quantum analogue of the classical momentum flux equation:

\begin{equation} \label{momentum flux classical}
    \frac{\partial}{\partial t} (\rho v_i)=-\frac{\partial p}{\partial x_i}-\frac{\partial}{\partial x_k} (\rho v_i v_k)
\end{equation}

See, e.g., \cite{LandauLifschitz}

As it is the case with classical hydrodynamics, conservation of total momentum is then expressed through the divergence theorem.

\subsection*{Energy Density Flux Equation}
In the Heisenberg representation the energy density flux equation is given by

\begin{equation} \label{Heisenberg-h}
    \partial_{t}\hat{h}(\mathbf{r}_{2})=\frac{i}{\hbar}\left[\hat{H},\hat{h}(\mathbf{r}_{2})\right]
\end{equation}

It can be shown that the result is given by 

\begin{equation} \label{energy-flux-result}
    \partial_{t}\hat{h}(\mathbf{r}_{2})=-\left(\begin{array}{c}
\left(\begin{array}{c}
\frac{1}{4}\left(\hat{v}_{a}\left(\mathbf{r}_{2}\right)\left[\frac{\partial}{\partial r_{a}}\hat{v}_{b}(\mathbf{r}_{2})\right]\hat{\rho}(\mathbf{r}_{2})\hat{v}_{b}\left(\mathbf{r}_{2}\right)+\left[\frac{\partial}{\partial r_{a}}\hat{v}_{b}(\mathbf{r}_{2})\right]\hat{v}_{a}\left(\mathbf{r}_{2}\right)\hat{\rho}(\mathbf{r}_{2})\hat{v}_{b}\left(\mathbf{r}_{2}\right)\right)\\
+\frac{1}{4}\left(\hat{v}_{b}\left(\mathbf{r}_{2}\right)\hat{\rho}(\mathbf{r}_{2})\hat{v}_{a}\left(\mathbf{r}_{2}\right)\left[\frac{\partial}{\partial r_{a}}\hat{v}_{b}(\mathbf{r}_{2})\right]+\hat{v}_{b}\left(\mathbf{r}_{2}\right)\left[\frac{\partial}{\partial r_{a}}\hat{v}_{b}(\mathbf{r}_{2})\right]\hat{\rho}(\mathbf{r}_{2})\hat{v}_{a}\left(\mathbf{r}_{2}\right)\right)\\
+\hat{v}_{b}\left(\mathbf{r}_{2}\right)\left[\frac{\partial}{\partial r_{a}}\hat{v}_{a}\left(\mathbf{r}_{2}\right)\right]\hat{\rho}(\mathbf{r}_{2})\hat{v}_{b}\left(\mathbf{r}_{2}\right)+\hat{v}_{b}\left(\mathbf{r}_{2}\right)\left[\frac{\partial}{\partial r_{a}}\hat{\rho}(\mathbf{r}_{2})\right]\hat{v}_{a}\left(\mathbf{r}_{2}\right)\hat{v}_{b}\left(\mathbf{r}_{2}\right)
\end{array}\right)\\
\\
+\frac{1}{2}\left(\begin{array}{c}
\left[\frac{\partial}{\partial r_{a}}\hat{v}_{a}\left(\mathbf{r}_{2}\right)\right]\left(\hat{\varepsilon}\left(\mathbf{r}_{2}\right)+\mathfrak{\hat{p}}\left(\mathbf{r}_{2}\right)\right)+\left(\hat{\varepsilon}\left(\mathbf{r}_{2}\right)+\mathfrak{\hat{p}}\left(\mathbf{r}_{2}\right)\right)\left[\frac{\partial}{\partial r_{a}}\hat{v}_{a}\left(\mathbf{r}_{2}\right)\right]\\
+\hat{v}_{a}\left(\mathbf{r}_{2}\right)\frac{\partial}{\partial r_{a}}\left(\hat{\varepsilon}(\hat{\rho}(\mathbf{r}_{2})+\mathfrak{\hat{p}}\left(\mathbf{r}_{2}\right)\right)\hat{v}_{a}\left(\mathbf{r}_{2}\right)
\end{array}\right)
\end{array}\right)
\end{equation}

Here, use has been made of the thermodynamic relation $\frac{\partial\varepsilon}{\partial\rho}=\mu=\frac{\varepsilon+p}{\rho}$.

Equation \eqref{energy-flux-result} is the symmetrized, quantum energy density flux equation of the classical energy density flux equation:

\begin{equation}\label{energy density flux classical}
    \frac{\partial}{\partial t} \bigl( \frac{\rho v^2}{2}+\varepsilon\bigr)=-\mathrm{div} \biggl( \mathbf v \bigl(\frac{\rho v^2}{2} + \left( \varepsilon + p\right) \bigr) \biggr)
\end{equation}

See e.g. \cite{LandauLifschitz}.

\subsection*{Entropy Density Flux Equation}
According to thermodynamics and statistical physics the entropy $s$ per unit mass at a given temperature $T$ is a constant in an ideal liquid . The entropy density operator can be therefore be defined as 

\begin{equation} \label{entropy-density-operator}
\begin{aligned}
    \hat{s}(\mathbf{r}_{1})&\equiv s\hat{\rho}(\mathbf{r}_{1})\\
     \left(\hat{s}(\mathbf{r}_{1})\right)^\dagger&= \hat{s}(\mathbf{r}_{1})
    \end{aligned}
\end{equation}

The corresponding flux equation expressing the conservation of entropy assumes the following guise

\begin{equation} \label{Heisenberg-s}
\begin{aligned}
    \partial_{t}\hat{s}(\mathbf{r}_{2})&=\frac{i}{\hbar}\left[\hat{H},\hat{s}(\mathbf{r}_{2})\right]\\
    &=-\frac{1}{2}\frac{\partial}{\partial r_{a}}\biggl(\hat{v}_{a}(\mathbf{r}_{2})\left(s\hat{\rho}(\mathbf{r}_{2})\right)+\left(s\hat{\rho}(\mathbf{r}_{2})\right)\hat{v}_{a}(\mathbf{r}_{2})\biggr)\\
    &\equiv -\mathrm{div}\hat{j}_{s}(\mathbf{r}_{2})
    \end{aligned}
\end{equation}
\\

Here we have defined the quantum analogue of the entropy current density $\hat j_s(\mathbf r_1)$ as 

\begin{equation} \label{entropy-current-density}
\begin{aligned}
    j_s(\mathbf r_1)&= \frac{1}{2}\biggl( \hat{v}_{a}(\mathbf{r}_{1})\left(s\hat{\rho}(\mathbf{r}_{1})\right)+\left(s\hat{\rho}(\mathbf{r}_{1})\right)\hat{v}_{a}(\mathbf{r}_{1}) \biggr)\\
    &=\frac{1}{2}\biggl( \hat{v}_{a}(\mathbf{r}_{1}) \hat{s}(\mathbf{r}_{1})+\hat{s}(\mathbf{r}_{1})\hat{v}_{a}(\mathbf{r}_{1}) \biggr)
    \end{aligned}
\end{equation}

Equation \eqref{Heisenberg-s} is the quantum analogue of the classical entropy current density balance equation:

 \begin{equation} \label{entropy flux classical}
     \frac{\partial(\rho s)}{\partial t}+\mathrm{div}(\rho s \mathbf v)=0
 \end{equation}

 See e.g. \cite{LandauLifschitz}.

Eq. \eqref{Heisenberg-s} merely means that the entropy $s$ is carried along with the mass density and that it is, of course, conserved. In the limit $T\rightarrow0$ then $s\rightarrow 0$.

\subsection*{Vorticity Flux Equation}

Defining the vorticity operator $\hat \omega_c (\mathbf r_1)$ as 

\begin{equation} \label{vorticity-operator}
    \begin{aligned}
        \hat{\omega}_{c}(\mathbf{r}_{1})&\equiv\left(\mathrm{rot}\,\hat{v}(\mathbf{r})\right)_{c}=\varepsilon_{abc}\frac{\partial}{\partial r}_{a}\hat{v}_{b}(\mathbf{r}_{1})\\
        \left(\hat{\omega}_{c}(\mathbf{r}_{1})\right)^{\dagger}&=\hat{\omega}_{c}(\mathbf{r}_{1})
    \end{aligned}
\end{equation}

the Heisenberg equation of motion of the vorticity operator is given by

\begin{equation} \label{Heisenberg-omega}
\begin{aligned}
    \partial_{t}\hat{\omega}_{c}(\mathbf{r}_{2})&=\frac{i}{\hbar}\left[\hat{H},\hat{\omega}_{c}(\mathbf{r}_{2})\right]
    \end{aligned}
\end{equation}

Evaluation of the commutator in \eqref{Heisenberg-omega} gives

\begin{equation} \label{Heisenberg-omega2}
\begin{aligned}
    \partial_{t}\hat{\omega}_{c}(\mathbf{r}_{2})=\frac{1}{2}\varepsilon_{bdc}\frac{\partial}{\partial r}_{b}\left(\hat{v}_{a}(\mathbf{r}_{2})\left(\frac{\partial v_{a}(\mathbf{r}_{2})}{\partial r_{d}}-\frac{\partial v_{d}(\mathbf{r}_{2})}{\partial r_{a}}\right)+\left(\frac{\partial v_{a}(\mathbf{r}_{2})}{\partial r_{d}}-\frac{\partial v_{d}(\mathbf{r}_{2})}{\partial r_{a}}\right)\hat{v}_{a}(\mathbf{r}_{2})\right)
    \end{aligned}
\end{equation}

 This, again, is the symmetric quantum analogue of the vorticity flux equation of classical hydrodynamics:
  \begin{equation} \label{vorticity flux classical}
     \frac{\partial(\mathrm{rot}\,\mathbf v)}{\partial t}=\mathrm{rot} \left(\mathbf v \times \mathrm{rot}\, \mathbf v \right)
 \end{equation}

 See e.g. \cite{LandauLifschitz}.

\subsection*{The Quantum Nature of the Vorticity}

In classical hydrodynamics, the circulation $\Gamma$ of the fluid velocity $\mathbf v$ is defined according to

\begin{equation} \label{Gamma classical}
    \Gamma=\oint \mathbf v \cdot d\mathbf l=\mathrm{const.}
\end{equation}
As it is well known, the circulation is conserved in time \cite{LandauLifschitz}.

In LQHD the velocity $\hat v_a (\mathbf r_1)$ is quantised by nature. One can sit at an arbitrary point $\mathbf{r}_{1}$ of observation and some fluid element will cross this point. The velocity at this point is given by $\hat {\mathbf v}=\frac{\mathbf {\hat p}}{m}$. The circulation at this point is given by 

\begin{equation} \label{Gamma quantum}
\begin{aligned}
    \Gamma&=\oint \mathbf {\hat v} \cdot d\mathbf l=\frac{1}{m}\oint\hat{p}_{a}dl_{a}=\frac{2\pi \hbar n}{m}\\
    n&\in\mathbb{N}
    \end{aligned}
\end{equation}

Consequently, in LQHD there also holds $\frac{d}{dt}\Gamma=\hat0$. As the point of observation $\mathbf{r}_{1}$ can be chosen arbitrarily, in the overall liquid the circulation $\Gamma$ is quantised in units of $\hbar$. Stokes Theorem will then confirm that the vorticity $\hat \omega_c (\mathbf r _1)$ is quantised, too.


\section{Discussion} 

\subsection{Conceptual Meaning}
\begin{figure}
    \centering
\includegraphics[width=0.55\linewidth]{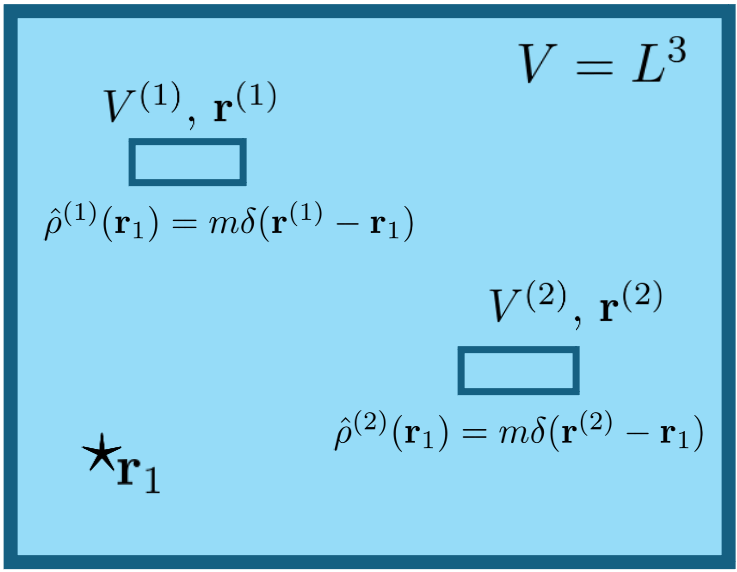}
    \caption{Bucket of Volume $V$ filled with an ideal quantum fluid decomposed into fluid elements $\alpha \in \{1,2\}$ of volume $V^{(\alpha)}$ at their positions $\mathbf r^{(\alpha)}$ and mass densities $\hat \rho^{(\alpha)} (\mathbf r_1)=m \delta(\mathbf r^{(\alpha)}-\mathbf r_1)$. The point of observation is at $\mathbf r _1$. }
    \label{fig:bucket}
\end{figure}

The equations of motion of the classical hydrodynamics of an ideal fluid can be constructed by considering a large volume $V=L^3$ filled with the fluid, which is then decomposed into smaller fluid elements $V^{(\alpha)}$. The decomposition is carried out such that 

\begin{equation}
    \sum_{\alpha=1}^{N^{(F)}} V^{(\alpha)}=V
\end{equation}
${N^{(F)}}$ is the number of fluid elements in the liquid \cite{LandauLifschitz}. This is sketched in fig. \ref{fig:bucket}. 

As the fluid is ideal, the fluid elements $\alpha$ do not interact with each other in a way that friction (shearing) or dissipation would occur.

If the length scales are chosen to be 

\begin{equation} \label{lengthscales HD}
    L^{\mathrm{(qm)}} \ll L^{(\alpha)} \ll L
\end{equation}

where $L^{\mathrm{(qm)}}$ is a typical length scale of the realms of quantum mechanics or atomic physics, say the Bohr radius $a_B=0.5 \AA$, such a single fluid element $\alpha$ of volume $V^{(\alpha)}=10^{-6}\, \mathrm{mm}^3$  contains $N^{(\alpha)}=10^{10}$ atoms or molecules. This follows from the ideal gas law. Hence, under such circumstances there also holds

\begin{equation} \label{particlenumber HD}
    N^{\mathrm{(qm)}}=1...\sim 10^3 \ll N^{(\alpha)}=10^{10} \ll N=10^{23}
\end{equation}

In this sense classical hydrodynamics is a mesoscopic theory. 

LQHD can be interpreted as quantum theory of a single--fluid element with the label $\alpha$ still containing $N^{(\alpha)}=10^{10}$ atoms or molecules. The role of a single particle described by quantum mechanics is then taken by a single fluid element described by LQHD. The fluid element cell, carrying $10^{10}$ atoms or molecules, obeys to the Schrödinger equation as 

\begin{equation}
    i\hbar \frac{\partial}{\partial t} \psi^{(\alpha)}(\mathbf r,t) =\hat H^{(\alpha)}(\mathbf r)\psi^{(\alpha)}(\mathbf r,t) 
\end{equation}

Hence, $\psi^{(\alpha)}(\mathbf r,t)=\exp\left(-\frac{i}{\hbar}\hat H^{(\alpha)} (\mathbf r)t\right) \varphi^{(\alpha)}(\mathbf r)$ is the wave function of fluid element $\mathrm{n^{o}}$ $\alpha$ and

\begin{equation} \label{h-alpha}
\begin{aligned}
    \hat H^{(\alpha)}(\mathbf r)=\int d^3 r_1 \left( \frac{\hat{v}^{(\alpha)}_{a} (\mathbf{r}_{1})\hat{\rho}^{(\alpha)}(\mathbf{r}_{1})\hat{v}^{(\alpha)}_{a}(\mathbf{r}_{1})}{2}+\hat{\varepsilon}^{(\alpha)}\left(\hat{\rho}^{(\alpha)}(\mathbf{r}_{1})\right)\right)
    \end{aligned}
\end{equation}

is the Hamiltonian of fluid element $\alpha$. In that case there holds

\begin{equation} \label{operators-alpha}
\begin{aligned}
\hat{\rho}^{(\alpha)}(\mathbf{r}_{1})&=m^{(\alpha)}\delta(\mathbf r^{(\alpha)}-\mathbf r_1)\\
\hat j_a^{(\alpha)}(\mathbf{r}_{1})&=\frac{1}{2} \left(\hat p_a^{(\alpha)} \delta(\mathbf r^{(\alpha)}-\mathbf r_1)+\delta(\mathbf r^{(\alpha)}-\mathbf r_1)\hat p_a^{(\alpha)} \right)\\
\hat v_a^{(\alpha)}(\mathbf{r}_{1})&=\frac{1}{2}\left(\frac{1}{\hat{\rho}^{(\alpha)}(\mathbf{r}_{1})}\hat j_a^{(\alpha)}(\mathbf{r}_{1}) +\hat j_a^{(\alpha)}(\mathbf{r}_{1})\frac{1}{\hat{\rho}^{(\alpha)}(\mathbf{r}_{1})} \right)
\end{aligned}
\end{equation}

As has been pointed out by Landau \cite{Landau41}, operators of different fluid element labels $\alpha,\beta$ commute with each other, so one can drop the index $\alpha$ and consider the evolution of a single fluid element only. 

It is then an interesting feature that the principles of quantum mechanics, e.g. the algebra of the commutators, may still be valid on the mesoscopic length scale of classical hydrodynamics and yield a complete theory of a quantum liquid (in the sense that in the classical limit of LQHD the equations of motion of classical hydrodynamics of an ideal, classical fluid are recovered). 

There is, however, one conceptual problem that may be more severe: the quantization of a hydrodynamic variable means that one carries out a re--quantization of a statistically averaged quantity. The internal energy $\varepsilon$ emerges from thermodynamics and statistical physics as

\begin{equation} \label{inner energy classical}
  \varepsilon= \langle \hat H \rangle_{gc}=\mathrm{Tr \left(\hat \rho_{gc} \circ \hat H \right)}
\end{equation}

$\hat \rho_{gc}$ is, e.g., the statistical operator of the grandcanonical ensemble.
Hence, speaking about the internal energy $\varepsilon$ seems to make sense only in a non quantum--statistical way.

It is thus quite possible that a re--quantization of such a thermodynamic quantity like the internal energy, e.g. introducing an internal energy operator $\hat \varepsilon \left(\hat \rho(\mathbf r_1)\right)$, is meaningless, and it may therefore very well be that, in the end, despite its allurement, LQHD will remain a beautiful, but coincidental phenomenon. On the other hand, within the length scales \eqref{lengthscales HD}, such a re-quantization of the internal energy may be reasonable. If there is (at least in part) a self--similarity in the laws of nature which can be applied to such a hierarchy of different length--scales \eqref{lengthscales HD} then the quantum average \eqref{inner energy classical} may be very well applied to a system of $10^{10}$ atoms or molecules within one fluid element. Single--atom quantum mechanics is then averaged out on the length scales $L^{\mathrm{(qm)}}$ (e.g. the Bohr radius). Owing to the fact that the quantum liquid itself consists of \textit{many} non--interacting fluid elements, the principles of quantum mechanics may be very well applicable on the mesoscopic length scale $L^{(\alpha)}$.

Such questions can only be decided by the experiment. If it is possible to bring LQHD to a form that can be used to make predictions for experiments, or that can be compared with predictions from other theories of superfluid Helium, it may become clear whether LQHD is at least useful.

The inner energy $\hat \varepsilon (\hat \rho (\mathbf r_1))$ is the variable which encodes what makes the liquid a liquid as it encodes the (non--dissipating) interactions between single atoms or molecules within one fluid element cell. The simplest description of liquids is given by real gases like the van--der Waals gas. One could therefore relate the pressure $\mathfrak{\hat{p}} (\mathbf r_1)$ of the quantum liquid to the van--der Waals pressure.

\subsection{Velocity Operator and Inverse of Mass Density Operator}

The introduction of the velocity operator 

\begin{equation} \label{velocity operator}
    \hat v_a(\mathbf r_1)\equiv \frac{1}{2}\left(\frac{1}{\hat{\rho}(\mathbf{r}_{1})}\hat{j}_{a}(\mathbf{r}_{1})+\hat{j}_{a}(\mathbf{r}_{1})\frac{1}{\hat{\rho}(\mathbf{r}_{1})}\right)
\end{equation}

is at the heart of LQHD. Please note that the velocity operator \eqref{velocity operator} has not the units of a density, but meters per second. 

There has been a discussion about whether the velocity operator \eqref{velocity operator} is a meaningful quantity or whether it is flawed by definition, and whether the momentum density operator $\hat j_a(\mathbf r_1)$ is the fundamental variable of the theory, or the velocity operator \cite{London45}. 

It has to be emphasised that, if one is interested in formulating a theory of a liquid in motion, such a question seems to be ill-defined, as for a hydrodynamic system, one will need the momentum density \textit{and} the velocity field.

What can be stated about the momentum density operator $\hat j_a(\mathbf r_1)$ and the velocity operator $\hat v_a (\mathbf r_1)$ within LQHD is first, the momentum density operator is the conjugated variable to the position operator $\hat r_b$:

\begin{equation} \label{commmutator-j,r}
    \begin{aligned}
        \left[\hat{j}_{a}(\mathbf{r}_{1}),\hat{r}_{b}\right]=\frac{\hbar}{i}\delta_{ab}\delta(\mathbf{r}-\mathbf{r}_{1})
    \end{aligned}
\end{equation}

The commutation relation between the velocity operator and the position operator is given by

\begin{equation} \label{commmutator-v,r}
\left[\hat{v}_{a}(\mathbf{r}_{1}),\hat{r}_{b}\right]=\frac{\hbar}{i}\frac{1}{\hat{\rho}(\mathbf{r}_{1})}\delta_{ab}\delta(\mathbf{r}-\mathbf{r}_{1})
\end{equation}

Second, the Heisenberg equation of motion of the position operator yields the velocity operator $\hat v_b=\frac{\hat p_b}{m}$ as

\begin{equation} \label{commmutator-j,r}
    \begin{aligned}
        \partial_{t}\hat{r}_{b}&=\frac{i}{\hbar}\left[\int d^{3}r_{1}\left(\frac{\hat{v}_{a}(\mathbf{r}_{1})\hat{\rho}(\mathbf{r}_{1})\hat{v}_{a}(\mathbf{r}_{1})}{2}+\hat{\varepsilon}\left(\hat{\rho}(\mathbf{r}_{1})\right)\right),\hat{r}_{b}\right]\\
        &=\int d^{3}r_{1}\hat{v}_{b}(\mathbf{r}_{1})\\
        &=\frac{\hat{p}_{b}}{m}\equiv \hat v_b
    \end{aligned}
\end{equation}

From that one can see that Landaus introduction of the velocity operator 
$\hat v_b (\mathbf r_1)$ is reasonable as it is constructed in accordance with the single--particle quantum physics limit of the momentum density operator $\hat j_b (\mathbf r_1)$ as the momentum operator $\hat{p}_{b}$, and it is in accordance with the classical limit of the momentum operator $\hat{p}_{b}$ as the momentum $p_a$.

As has been pointed out, without the velocity operator \eqref{velocity operator} there will not be a quantum hydrodynamic theory of a liquid, which is why it is not only a useful and meaningful concept, but very essential to the theory and thus remains a mark of Landau's great intuition and genius.

Another aspect which has been frequently criticised about LQHD, and which has to be introduced along with the velocity operator, is the usage of the inverse operator $\frac{1}{\hat \rho (\mathbf r_1)}$. Throughout all calculations this is, however, never a problem, as such expressions always cancel. If, though, in any calculation there would occur a gradient of $\frac{1}{\hat \rho (\mathbf r_1)}$ it would be given by 

\begin{equation}\label{gradient-inverse-rho}
    \left[\hat p_a, \frac{1}{\hat \rho (\mathbf r_1)}\right]=-\frac{1}{\left(\hat \rho (\mathbf r_1)\right)^2}\frac{\partial \delta (\mathbf r-\mathbf r_1)}{\partial r_a }
\end{equation}

which can be handled in the usual way (integration by parts etc.). Hence, introducing the inverse of the mass density operator into the theory is essential as it yields the correct density flux equations, and, from a formal point of view, is never problematic.

\subsection{Comparison of LQHD to the Two--Fluid Model and to the Gross--Pitaveskii Theory}

The hydrodynamic, purely classical, "Two--Fluid Model" (TFM) is well known to be \textit{the} dobbin of research on superfluid Helium 4. It is the phenomenological model for describing most of the outcomes of experiments. Within this model (or rather its many extensions) the liquid is described by two "interpenetrating" components, a normal fluid flow $\mathbf v_n$ and a superfluid fluid flow $\mathbf v_s$. In its most comprehensive formulation, the TFM describes dissipative processes at finite temperatures. For a good overview see, for example, \cite{Nemirovskii13}. 

However, as LQHD is a theory which, in its current stage, cannot take into account dissipation, and is formulated for $T=0$, the comparison must be made with the basic formulation of the TFM, which is in the zero--temperature limit and without dissipation. 

The assumption that two different flow fields are present in the liquid enables to derive from the Galilean invariance principle and the usual conservation laws of hydrodynamics the following system of flux equations

\begin{equation} \label{TFM}
\begin{aligned}
    0&=\partial_t \rho +\mathrm{div} \,\mathbf j\\
    0&=\partial_t  j_i+\frac{\partial \Pi_{ik}}{\partial r_k} \\
     0&=\partial_t \left(\rho s \right)+ \mathrm{div} \left(\rho s \mathbf v_n\right)\\
    0&=\partial_t \mathbf v_s+ \nabla \left(\frac{\mathbf v_s \mathbf v_s }{2} +\mu  \right)\\
    0&=\partial_t E+ \mathrm{div} \,\mathbf Q\\
   \mathbf j &=\rho_s \mathbf v_s + \rho_n \mathbf v_n \\
   \Pi_{ik}&=\rho_n v_{n,i}v_{n,k}+\rho_s v_{s,i}v_{s,k}+ p\delta_{ik}\\
   \mathbf Q&= \left( \mu +\frac{v_s v_s}{2}\right)\mathbf j + T \rho s \mathbf v_n + \rho_n\mathbf v_n (\mathbf v_n,\mathbf v_n -\mathbf v_s)
\end{aligned} 
\end{equation}

See e.g. \cite{LandauLifschitz}.

At $T=0$ there holds $\mathbf v_n=0$ and $s=0$. The set of equations \eqref{TFM} in this limit assumes the following guise

\begin{equation} \label{TFM zero temperature}
\begin{aligned}
    0&=\partial_t \rho +\mathrm{div} \,\left(\rho \mathbf v_s \right)\\
    0&=\partial_t  j_i+\frac{\partial \Pi_{ik}}{\partial r_k}\\
    0&=\partial_t \mathbf v_s+ \nabla \left(\frac{\mathbf v_s \mathbf v_s }{2} +\mu  \right)\\
    0&=\partial_t E+ \mathrm{div}\, \mathbf Q\\
    \mathbf j &=\rho_s \mathbf v_s \\
      \Pi_{ik}&=\rho_s v_{s,i}v_{s,k}+ p\delta_{ik}\\
      \mathbf Q&= \left( \mu +\frac{v_s v_s}{2}\right)\mathbf j 
\end{aligned} 
\end{equation}

Comparing the set of equations \eqref{TFM zero temperature} with \eqref{massbalance}, \eqref{momentum flux result} and \eqref{energy-flux-result}, respectively, it can be seen that even at $T=0$ the equations of motion of the TFM and LQHD are not equivalent. This is because in the frame of the TFM the velocity $\mathbf v_s$ is a \textit{classical} field, whereas within the frame of LQHD the equations of motion emerge in a symmetrized fashion because the densities themselves are quantised fields.

Next, for introducing a vorticity $\omega=\mathrm{rot} \,\mathbf v_s$ of the flow field into the TFM, the superfluid velocity component $\mathbf v_s$ has to be put into relation to the Gross--Pitaevskii theory (GPT), otherwise there is no formal argument as to why the circulation $\Gamma$ should be quantised.

The Gross-Pitaevskii theory is the simplest theory of a \textit{weakly interacting, dilute Bose--Einstein condensate (BEC)} at $T=0$. As it is derived from a many--particle Hamiltonian in second quantisation, it is a general theory which has to be interpreted when being applied to a system. In the case of superfluid Helium 4, the GPT is interpreted such that it describes the dynamics of the density of the Bose--particles in the ground state. Excitations out of this ground state - e.g. when one condensed atom is kicked out or added to it - are related to the creation and annihilation of elementary excitations - phonons of all kinds. The phonons can interact with each other and have their own dynamics next to the dynamics of the Bose--condensed atoms in the quantum gas. In that sense superfluid Helium is described as a depleted Bose--Einstein condensate where the depletion is coming from the interactions between the condensed Bosons.

In the Bogoliubov approximation of the GPT (the field operators are assumed to commute because the number of condensed atoms is very large compared to a single atom being added to, or kicked out of the ground state), and for weak interactions (the atoms interact via the "contact interaction" which is a delta--like interaction), the time--dependent Gross Pitaevskii equation (GPE) assumes the form of a "nonlinear Schrödinger equation", 

\begin{equation} \label{GPE}
\begin{aligned}
    i\hbar \partial_t \phi(\mathbf r,t)&= \left(\frac{\hat{\mathbf p}^2}{2m}+g |\phi(\mathbf r,t)|^2 \right)\phi(\mathbf r,t)\\
    g&\equiv\frac{4\pi \hbar^2}{m}a   \\
    \phi(\mathbf r,t)&=\sqrt{n(\mathbf r,t)} \exp\left( i\varphi (\mathbf r,t)\right)
    \end{aligned}
\end{equation}

where $m$ is the mass of a Helium atom, $a$ is the scattering length of the contact interaction, $\sqrt{n(\mathbf r,t)}$ is the density of Bose condensed atoms, and the "macroscopic wave--function" $\phi(\mathbf r,t)$ now describes all condensed atoms collectively. 

Expanding the GPE \eqref{GPE} yields a set of hydrodynamic--like equations for the density $\rho\equiv n$ and the velocity $\mathbf v\equiv \frac{\hbar}{m}\nabla \varphi (r,t)$. This can be interpreted in such a way that a Bose--Einstein condensate can be considered as an ideal liquid, because the equation of motion of the velocity $\mathbf v$ has a similar form of the (classical) Euler equation:

\begin{equation} \label{Eulereq GPT}
 m\partial_t \mathbf v + \nabla \left( \frac{mv^2}{2}+ g n(\mathbf r,t) -\frac{\hbar^2}{2m} \sqrt{n(\mathbf r,t)} \nabla^2 n(\mathbf{r},t) \right) =\mathbf 0  
\end{equation}

The last term in \eqref{Eulereq GPT} is the so--called quantum pressure which is not present in the classical hydrodynamics of an ideal liquid.

As the velocity $\mathbf v$ of the Bosons in the ground state is, by definition, the gradient of a scalar function $\varphi (\mathbf r,t)$, it is irrotational. As this scalar function is the phase of the macroscopic wave--function, it must be single-valued. This gives then then quantisation of the circulation

\begin{equation}
    \Gamma= \oint \mathbf v \cdot d\mathbf l=\frac{\hbar}{m}\oint \nabla \varphi(\mathbf r,t)=\frac{ 2\pi \hbar n}{m}
\end{equation}

The equating of the superfluid velocity of the TFM and the velocity of the Bose--condensed particles $\mathbf v_s=\mathbf v$ is somewhat artificial. It would therefore be more satisfactory if one would have a quantum hydrodynamic theory in which the depleted Bose--condensation occurs in a more natural way. Also, that has been pointed out elsewhere (e.g. \cite{NozieresBook, LeggettBook, GriffinBook}), a liquid is manifestly not a dilute system of weakly interacting particles.

LQHD is a quantum theory of the ideal liquid itself (not a theory of elementary excitations), due to the presence of the velocity operator $\hat {\mathbf v}_a (\mathbf r_1)$ and the internal energy density operator $\hat \varepsilon (\hat\rho ))$. In a quantum liquid described by LQHD $\mathrm{curl}\, \hat {\mathbf v}( \mathbf r_1)\neq \hat0 $ is nowhere zero and is at minimum of the order of $\hbar$. $\hat v_s (\mathbf r_1)$ is not necessarily a potential flow and therefore much more general, and the vorticity and the circulation $\Gamma$ is quantised by the very nature of the theory.

\section{Summary}

As has been argumented, LQHD describes the \textit{liquid itself} in its quantum nature. Building up a fundamental quantum theory of the liquid may henceforth provide a complementary approach in the theoretical research of superfluid Helium which to date has mainly dealt with research on elementary excitations (phonons of all kinds) in the liquid \cite{GriffinBook}. The basis of such a quantum theory of a liquid has been given by Landau in 1941 who has shown that, by introducing the velocity operator and the internal energy density operator, and by applying the usual requirements and formal rules of quantum mechanics together with a similar rule as the operation \textcolor{magenta}{\FiveFlowerOpen}, the quantum analogue of the mass density flux equation and the quantum Euler equation emerges. The current work shows that, by applying the operations \textcolor{magenta}{\FiveFlowerOpen} and \textcolor{green}{\EightFlowerPetal} proposed in sec. \ref{sec:Methods}, one can go further and derive a quantum analogue of the momentum density flux, the energy density flux, the vorticity flux and the entropy density flux. The latter may be used to formulate LQHD for finite temperatures. 

It was also shown that the velocity operator is an essential ingredient of the theory, and that it is in agreement with the classical limit of the momentum density operator. Next, the usage of the inverse of the mass density operator is essential, too, and does, at no point, provide any problem of formal nor conceptual kind. 

In its current form, LQHD is the simplest theory of a quantum \textit{liquid}. The next step is the formulation of LQHD for finite temperatures $T$ for which the internal energy density operator becomes a function of the entropy density $\hat s(\mathbf r_1)$ as $\hat \varepsilon(\hat \rho))\longrightarrow \hat \varepsilon(\hat \rho,\hat s))$. 

Then it is aimed at studying the quantum liquid under rotation. For this one has to define the angular velocity operator and extend the Hamiltonian \eqref{Hamiltonian} by terms representing the centrifugal force and the Coriolis force. 

An extension to a full quantum theory of the Navier--Stokes equation is highly eligible, as such a theory may provide a fundamental quantum theory of the Two--Fluid--Model or Quantum Turbulence. Building up such a quantum theory of dissipative liquids requires to find a term in the Hamiltonian \eqref{Hamiltonian} which produces the Laplacian of the velocity $\hat v_a (\mathbf r_1)$ in the Heisenberg equation of motion, and, consistently, the viscous contribution (the Cauchy stress tensor) to the momentum density operator. If such a proceeding is possible and meaningful is left to future research.

\textbf{Acknowledgements}\\
The author gives thanks to Michal Pavelka, Emil Varga and Grigory Volovik for reading through the manuscript and for useful discussions. This work is supported by the Czech Science Foundation (grant number 23-05736S).



\end{document}